\documentclass[journal=ascefj,manuscript=article,layout=twocolumn]{achemso}

\makeatletter
\let\l@addto@macro\relax
\makeatother
\usepackage[fontsize=9pt]{scrextend}
\usepackage{chemformula} 
\usepackage[T1]{fontenc} 
\pdfoutput=1

\SectionsOn

\author{P. Thapa}
\email{pawan.thapa@uta.edu}
\affiliation{Department of Physics, University of Texas at Arlington, Arlington, TX 76019}
\alsoaffiliation{Department of Chemistry and Biochemistry, University of Texas at Arlington, Arlington, TX 76019} 
\author{N. K. Byrnes}
\email{byrnes.nicholas@mavs.uta.edu}
\affiliation{Department of Physics, University of Texas at Arlington, Arlington, TX 76019}\author{A. A. Denisenko}
\author{F.W. Foss, Jr.}
\affiliation{Department of Chemistry and Biochemistry, University of Texas at Arlington, Arlington, TX 76019}
\author{B.J.P. Jones}
\affiliation{Department of Physics, University of Texas at Arlington, Arlington, TX 76019}
\author{J. X. Mao}
\affiliation{Department of Chemistry and Biochemistry, University of Texas at Arlington, Arlington, TX 76019}
\author{K. Nam}
\affiliation{Department of Chemistry and Biochemistry, University of Texas at Arlington, Arlington, TX 76019}
\author{C.A. Newhouse}
\affiliation{Department of Chemistry and Biochemistry, University of Texas at Arlington, Arlington, TX 76019}
\author{D.R. Nygren}
\affiliation{Department of Physics, University of Texas at Arlington, Arlington, TX 76019}
\author{A.D. McDonald}
\affiliation{Department of Physics, University of Texas at Arlington, Arlington, TX 76019}
\author{T.T. Vuong}
\affiliation{Department of Chemistry and Biochemistry, University of Texas at Arlington, Arlington, TX 76019}
\author{K. Woodruff}
\affiliation{Department of Physics, University of Texas at Arlington, Arlington, TX 76019}

\title{Barium Selective Chemosensing by Diazacrown Ether Naphthalimide Turn-on Fluorophores for Single Ion Barium Tagging}

\abbreviations{IR,NMR,UV}

\begin{document}

\begin{abstract}
   Single molecule fluorescence detection of barium is investigated for enhancing the sensitivity and robustness of a neutrinoless double beta decay ($0\nu\beta\beta$) search in $^{136}$Xe, the discovery of which would alter our understanding of the nature of neutrinos and the early history of the Universe. A key developmental step is the synthesis of barium-selective chemosensors capable of incorporation into ongoing experiments in high-pressure $^{136}$Xe gas.  Here we report turn-on fluorescent naphthalimide chemosensors containing monoaza- and diaza-crown ethers as agents for single Ba$^{2+}$ detection.  Monoaza-18-crown-6 ether naphthalimide sensors showed sensitivity primarily to Ba$^{2+}$ and Hg$^{2+}$, whereas two diaza-18-crown-6 ether naphthalimides revealed a desirable selectivity toward Ba$^{2+}$. Solution-phase fluorescence and NMR experiments support a photoinduced electron transfer mechanism enabling turn-on fluorescence sensing in the presence of barium ions. Changes in ion-receptor interactions enable effective selectivity between competitive barium, mercury, and potassium ions, with detailed calculations correctly predicting fluorescence responses.  With these molecules, dry-phase single Ba$^{2+}$ ion imaging with turn-on fluorescence is realized using oil-free microscopy techniques.  This represents a significant advance toward a practical method of single Ba$^{2+}$ detection within large volumes of $^{136}$Xe, plausibly enabling a background-free technique to search for the hypothetical process of $0\nu\beta\beta$.
\end{abstract}

\section{Introduction}
The development of single barium ion sensing techniques in gaseous or liquid xenon environments is known as ``barium tagging''~\cite{Moe:1991ik} and is a vibrant area of R\&D in nuclear physics~\cite{Chambers:2018srx,mcdonald2018demonstration}. If realized with high efficiency and at large scales, dry detection of individual Ba$^{2+}$ in coincidence with well-measured ionization charge deposits in xenon would greatly improve discrimination against relatively copious background processes due to radioactivity, and would enhance discovery reach of searches for the hypothetical radioactive process of $0\nu\beta\beta$.  A robust observation would demonstrate that the neutrino is its own antiparticle, a discovery with major impact for particle and nuclear physics as well as for cosmology.~\cite{Fukugita:1986hr}  One possible technical implementation of barium tagging employs single molecule fluorescent imaging (SMFI) with chemosensors that exhibit turn-on fluorescence in response to chelation with barium~\cite{nygren2016detection,jones2016single}.

Fluorescence-based chemosensors are sensitive analytical tools for rapid chemical and biochemical measurement.~\cite{de1997signaling,valeur2000design,hamilton2015optical} Properly incorporated within SMFI experiments, they reveal individual molecular events and fundamentals whose significance is unavailable through bulk analyses. 1,8-Naphthalimide fluorophores exhibit many favorable properties for developing new chemosensors.~\cite{duke2010colorimetric,panchenko2014fluorescent} Synthetically tractable 1,8-naphthalimides allow construction of both photoinduced electron transfer (PET) and intramolecular charge transfer (ICT) fluorescent sensors.~\cite{panchenko2013comparative}. Relatively large Stokes shifts and high photostability features in these fluorophores are beneficial for high signal-to-noise ratios.~\cite{gao2017fluorescent} Crown ethers afford enhanced efficiency in binding to metal ions and can act as switches for PET and ICT based sensors.\cite{alfimov1999fluorescence} As such, 1,8-naphthalimide fluorophores appended with crown ethers have been used as colorimetric and fluorescent sensors for the detection of ions and small molecules that are of great significance in environment, health, and energy~\cite{hou2011rapid,panchenko2015cation,panchenko2018selective,panchenko2019chemoselective}. Several 1,8-naphthalimide chemosensors have been developed for detection of cations, anions, and small polar molecules.\cite{georgiev2016synthesis,fernandez2017solid,aderinto2017synthesis,georgiev2019ratiometric} However, 1,8-naphthalimide based chemosensors with selective response to Ba$^{2+}$ remain underdeveloped.\cite{licchelli2006prototype,hamilton2015optical}


Selective barium sensing has been demonstrated by a relatively small number of instructive fluorescent chemosensors.\cite{patnaik2003handbook,zhao2010complexation,saluja2011benzthiazole,banerjee2011competitive,basa2011site,guo2013colorimetric,hamilton2015optical,garcia2017selective,chaichana2019selective,li2020simple} Barium sensing is not widely studied in clinical settings. Though acute barium exposure can be deadly, diagnoses are rare and often confirmed by inductively-coupled plasma mass spectrometry analysis of patient plasma and urine.\cite{kravchenko2014review,lukasik2014barium} Beyond the primary goals of barium tagging in $0\nu\beta\beta$, sensitive chemosensors for direct detection of Ba$^{2+}$ would be a promising clinical advance, and some recent reports exist. Nakahara and coworkers pioneered a monoazacryptand receptor with PET switch to pyrene fluorescence that operates in aqueous micellar systems.~\cite{nakahara2004novel,nakahara2005fluorometric} Crown ether systems invoke termolecular complexes with two crown moieties sandwiched around one Ba$^{2+}$.~\cite{licchelli2006prototype,kondo2011synthesis} A bioinspired G-quadruplex chemosensor provided a rapid detection of Ba$^{2+}$.~\cite{yang2013highly,xu2017highly} Recently, a fluorescent chemosensor based on phenoxazine system with an ICT turn-on mechanism was reported for Ba$^{2+}$, where charge transfer was acheived by metal binding specifically to a single amide functional group.~\cite{ravichandiran2019simple} These receptors are instructive, but unsuitable for dry-phase SMFI device design. 

Within the context of R\&D toward a barium-tagging phase of the NEXT program~\cite{monrabal2018next,martin2016sensitivity,adams2020sensitivity}, we previously demonstrated single ion sensitivity to Ba$^{2+}$ using commercial chemosensors in an aqueous suspension~\cite{mcdonald2018demonstration}.  Achieving single ion sensitivity mandated the use of a competitive binding agent BAPTA, and used fluorophores incompatible with dry operation.  To address the challenge of solventless sensing of barium, a class of dry-phase active fluorescent chemosensors using monoazacrown receptors was demonstrated.~\cite{thapa2019barium} Single-molecule sensitivity of this system described in Ref.~\cite{thapa2019barium}  was elusive due to UV excitation of impurities in the substrates and lack of photobleaching transitions to use for single molecule identification.   
Similar molecules have also recently been shown to be suitable for ratiometric or ``bi-color'' Ba$^{2+}$ sensing~\cite{Rivilla:2019vzd}, a promising new direction under consideration for barium tagging. 

\begin{figure}
\centering
\includegraphics[width=0.5\linewidth]{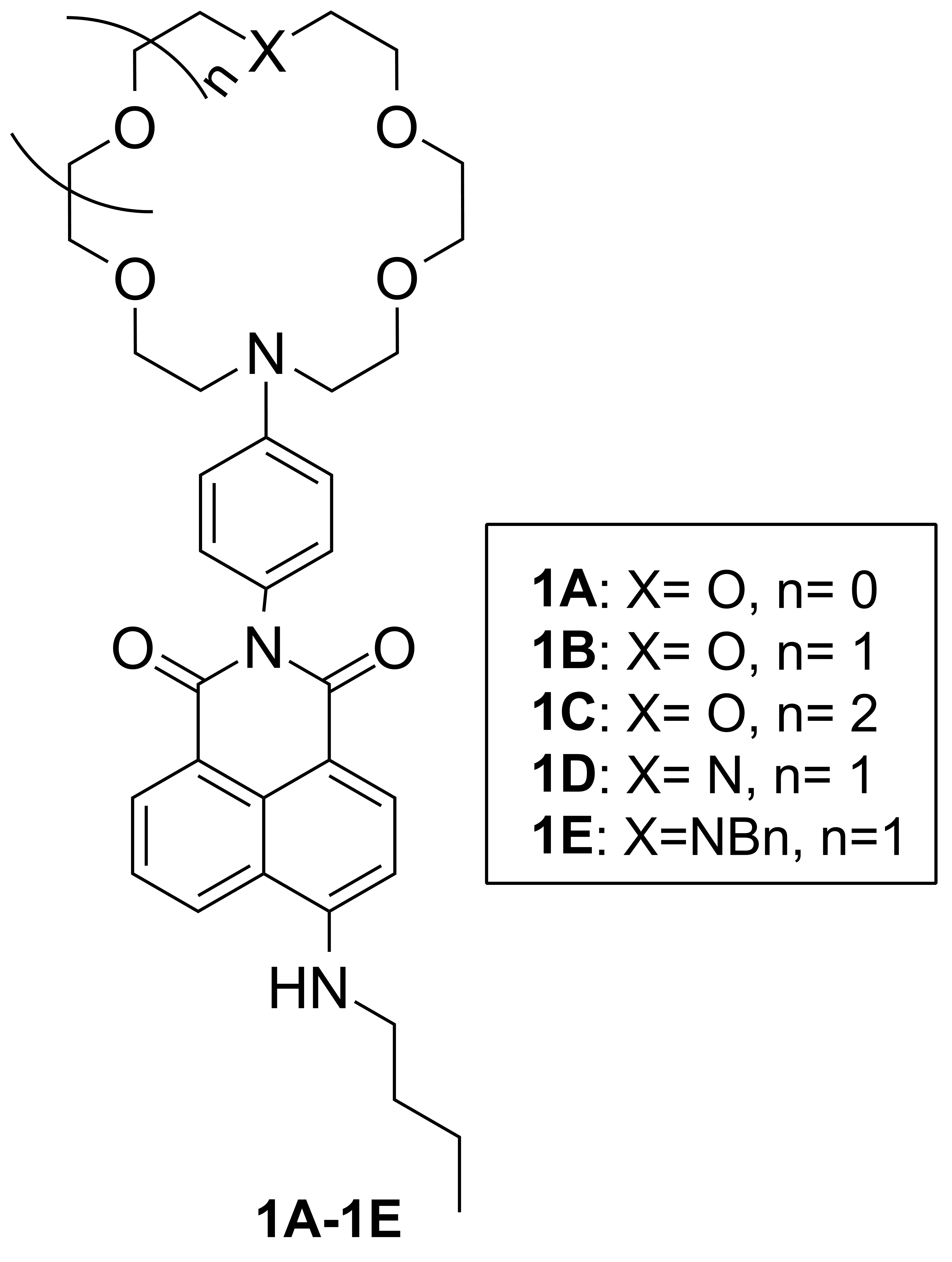}
\caption{Napthalimide derivatives used in this study}
\label{fig:NapStructure}
\end{figure}

In this paper we report a study on the synthesis, characterization, analysis, and computational modeling of a new family of monoazacrown and diazacrown naphthalimide fluorescent sensors (\textbf{Fig.~\ref{fig:NapStructure}})  with enhanced selectivity and sensitivity to Ba$^{2+}$. These visible fluorescent probes are amenable to SMFI microscopy in dry environments and consequently allowed us to resolve individual Ba$^{2+}$ ions without a competitive binding agent.  This extends the functionality of the dyes described in Ref.~\cite{thapa2019barium}, with visible excitation proving to be a critical ingredient in overcoming fluorescent backgrounds of glass and quartz substrates at the level required for single molecule sensitivity.   We also report on computational models that offer reliable predictability of target ion- and binding-site dependent fluorescent response within this family, a capability that will accelerate explorations of structure-function relationships in dry-phase SMFI probes.



\section{Experimental Section}

\subsection{Synthesis}
The synthesis of naphthalimide derivatives \textbf{1A-1E} is shown in \textbf{Scheme 1}. Monoazacrown ethers \textbf{2A-2C} were assembled as previously described.~\cite{luk2012synthesis,thapa2019barium} Diaza-18-crown-6 ethers were prepared by the modification of methods from a previous report.~\cite{koonrugsa2019metal} Complete details and chemical characterization can be found in the supplementary information.

\subsection{Sample preparation for solution-phase studies}
Stock solutions of probe molecules were prepared at 1$\times$10$^{-3}$~M concentration in acetonitrile solvent. Stock solutions of metal perchlorates were made at 1$\times$10$^{-2}$~M concentration in 9:1 acetonitrile/water mixture. UV-Vis studies were performed with and without Ba$^{2+}$ in acetonitrile, dichloromethane, acetone, and ethanol solvent with probe \textbf{1B} (10$\times$10$^{-6}$~M) and barium perchlorate (50$\times$10$^{-6}$~M).  Fluorescence titration and metal ion selectivity studies were performed by maintaining a final concentration of probe molecule at 1$\times$10$^{-6}$~M concentration. Metal ion selectivity studies were conducted by maintaining metal perchlorate concentration at 5$\times$10$^{-6}$~M. Competitive fluorescence experiments were performed using 2$\times$10$^{-6}$~M of the probe \textbf{1D} with 5$\times$10$^{-6}$~M metal perchlorate solution in acetonitrile solvent. Job's plot experiments were performed by maintaining a total concentration of the probe and barium perchlorate solution at 20$\times$10$^{-6}$~M. Samples were incubated for two minutes in the dark before measuring fluorescence intensity. $^1$H NMR titration experiments were conducted with the concentration of probe at 1$\times$10$^{-2}$~M and metal perchlorate (barium, potassium, and mercury) at 1$\times$10$^{-2}$~M and 5$\times$10$^{-2}$~M in acetonitrile-\textit{d}$_3$ solvent at room temperature. 

\subsection{Slide preparation for dry-phase fluorescence}
Dry samples of probe \textbf{1D} on glass slides for dry-phase fluorescence studies were prepared as follows: An acetonitrile solution of barium selective probe \textbf{1D} was mixed into cyanoacrylate glue to desired concentration (10$\times$10$^{-12}$~M to 10$\times$10$^{-6}$~M). 30~$\mu$L of this solution was applied to a glass microscope slide and then dried in an oven for five minutes at 373 K to set into a thick solid layer. Fluorescence response from three slides was measured with excitation at 430~nm, before and after washing with a 50~$\mu$L  of 1$\times$10$^{-3}$~M barium perchlorate in acetonitrile and then drying. Each slide was then re-scanned three times, and the fluorescence response was averaged to yield a mean response.

\subsection{Microscopy details}
The probe \textbf{1D} was suspended at a concentration of 10$^{-11}$~M within the cyanoacrylate matrix and analyzed via fluorescence microscopy. Excitation is delivered by a Supercontinuum laser with output selected by acousto-optical tunable filter. For the studies performed here, several output lines were superposed between 420 nm and 437.5 nm, to maximize output power and cover the excitation band of \textbf{1D}. Laser light was cleaned by passing through a beam-expander and cutting off the outer, non-uniform edge with an iris, generating a circular and approximately Gaussian beam profile. The beam was then redirected through a 500~nm short-pass filter to reduce incident background light, reflected off a 505~nm dichroic mirror, and focused through a 100$\times$ air-coupled microscope objective (NA=0.95) for a final power output of 125~mW to excite the \textbf{1D} matrix in epifluorescent mode. The objective collected the resultant Stoke's shifted fluorescent emissions, which were then passed back through the dichroic mirror and a 500~nm long-pass filter, again for background reduction, and collected via a Hamamatsu EMCCD camera, which acquired one image every half-second into 300-image sequences. 

Microscope images were processed using an algorithm that first filters the image in Fourier space to remove slowly sloping backgrounds and then sums the sequence over the full imaging time. After filtering, single molecule candidates can be identified after subtracting the background profile, which is found by blurring the summed image via a Gaussian filter. Single molecule candidates are identified as points with intensities in excess of $3\sigma$ above the background. Once the candidates are identified, their locations on the raw images are analyzed as a function of time for the duration of the sequence to produce fluorescent trajectories. These trajectories are scanned for an instantaneous drop or ``single step'' profile, which is the hallmark characteristic of a single molecule undergoing a discrete photo-bleaching process.

\subsection{Computation details}
Molecular structures were optimized using M06-2X\cite{zhao2008m06} functional with the SDD\cite{dolg1987energy,kaupp1991pseudopotential,bergner1993ab,dolg1993relativistic} effective core potential basis set for heavy metal atoms (barium, mercury) and def2-SVP\cite{weigend2005balanced,weigend2006accurate} basis set for other atoms, followed by frequency analysis to confirm the nature of their energy minima (no imaginary frequency). Calculations with time-dependent density-functional theory (TDDFT)\cite{runge1984density} were carried out to reveal the 
orbitals involved in the observed fluorescence events. SMD\cite{marenich2009universal} or PCM\cite{hall1995combined} solvation models were used to incorporate solvent effects with acetonitrile as the solvent. All calculations were performed using the Gaussian 09 program.~\cite{gaussian0920091}

\section{Results and discussions:}

\subsection{Synthesis}
\textbf{Scheme 1} represents the synthesis of probes \textbf{1A}-\textbf{1E}. In brief, nucleophilic aromatic substitution between 4-fluoronitrobenzene and mono/diaza crown ethers, prepared by modification of our previous work,\cite{thapa2019barium} gave the desired 4-nitroaniline derivatives \textbf{3A-3D}. Reducing 4-fluoronitrobenzene to one equivalent was critical in the case of diaza-18-crown-6 ether \textbf{2D}. Nitro reduction by Pd-catalyzed hydrogenolysis, followed by condensation with commercially available 4-bromonaphthalic anhydride resulted in the desired bromonaphthalimide products \textbf{4A-4D}. Initial attempts for the final nucleophilic aromatic substitution reaction of bromonaphthalimide with \textit{n}-butylamine in 2-methoxyethanol solvent led to poor yields of the desired aminonaphthalimides. However, \textit{N}-methyl-2-pyrrolidone (NMP) acting as a basic solvent enabled the bromide substitution to afford all desired aminonaphthalimides \textbf{1A-1D}. Compound \textbf{1E} was prepared by nucleophilic substitution by dropwise addition of  benzyl bromide into a solution containing \textbf{1D} and potassium carbonate. Overall, this synthetic approach led to the efficient preparation of the desired naphthalimide fluorescent probes \textbf{1A-1E}.

\begin{scheme}[t]
\label{Scheme1:Synthtic scheme}
\includegraphics[width=0.99\linewidth]{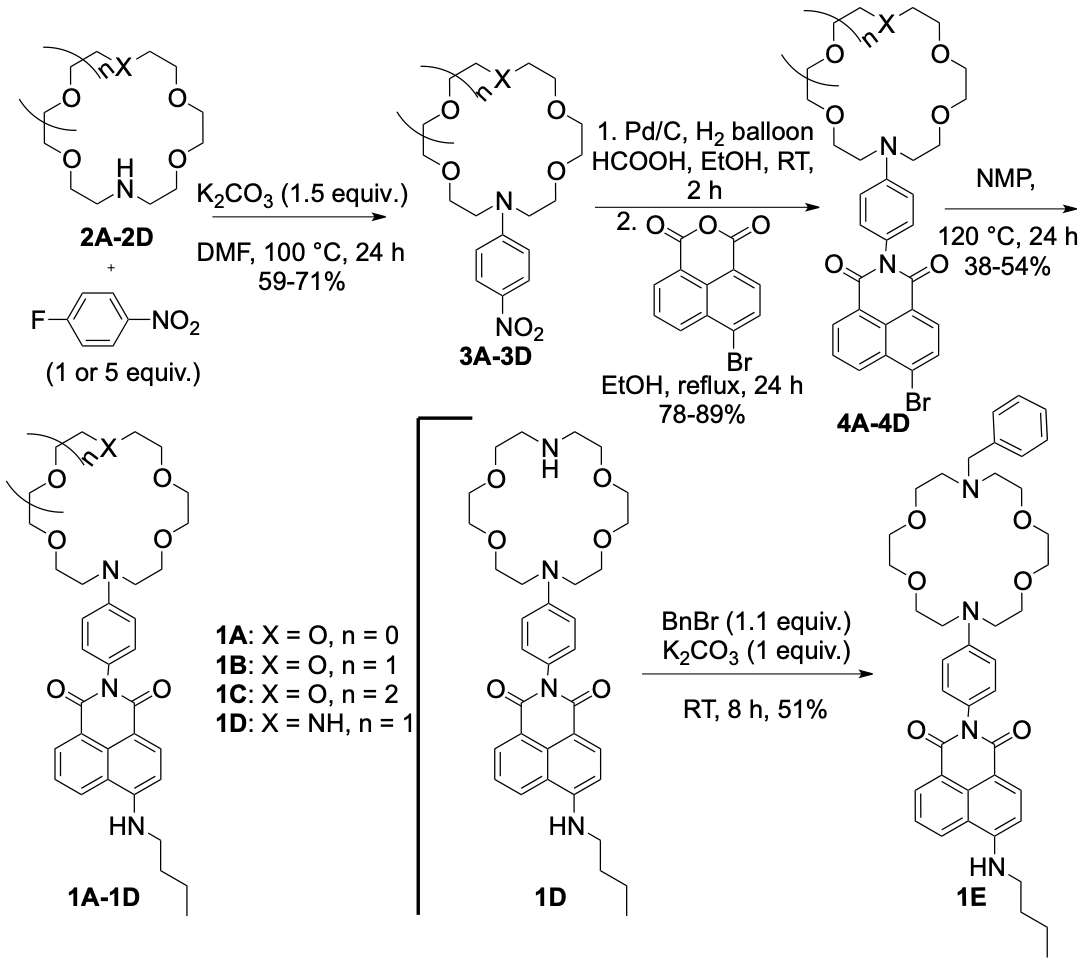}
\caption{Synthesis of 4-amino-1,8-naphthalimide sensors.}
\end{scheme}

\subsection{UV-Vis studies}
UV-Vis spectra of all synthesized 1,8-naphthalimide based fluorescent probes (\textbf{1A-1E}) were recorded in acetonitrile solvent at room temperature. Major absorbances were centered between 200-300 nm and 380-480 nm---consistent with 4-amino-1,8-naphthalimide chromophores~\cite{panchenko2014fluorescent}. Solutions doped with Ba$^{2+}$ provided no or negligible shifts on UV-Vis absorbances in all probes (\textbf{SI Fig. S2}). These effects were also encountered in dichloromethane, acetone, and ethanol (\textbf{SI Fig. S3}). Ba(ClO$_4$)$_2$ addition led to a concentration dependent increase in fluorescence centered near 530 nm in probes \textbf{1B}, \textbf{1D}, and \textbf{1E}, when excited at $\lambda_{max}$ obtained from UV-Vis spectrum, suggesting turn-on fluorescence by changes to photoinduced electron transfer (PET).~\cite{panchenko2018selective} Similarly, fluorescence analysis as a function of absorption wavelength from 430-460 nm showed no significant change in the emission intensity profile (\textbf{SI Fig. S4}).

\subsection{Fluorescence response and metal ion selectivity}
\begin{figure}[t]
\centering
\includegraphics[width=0.99\linewidth]{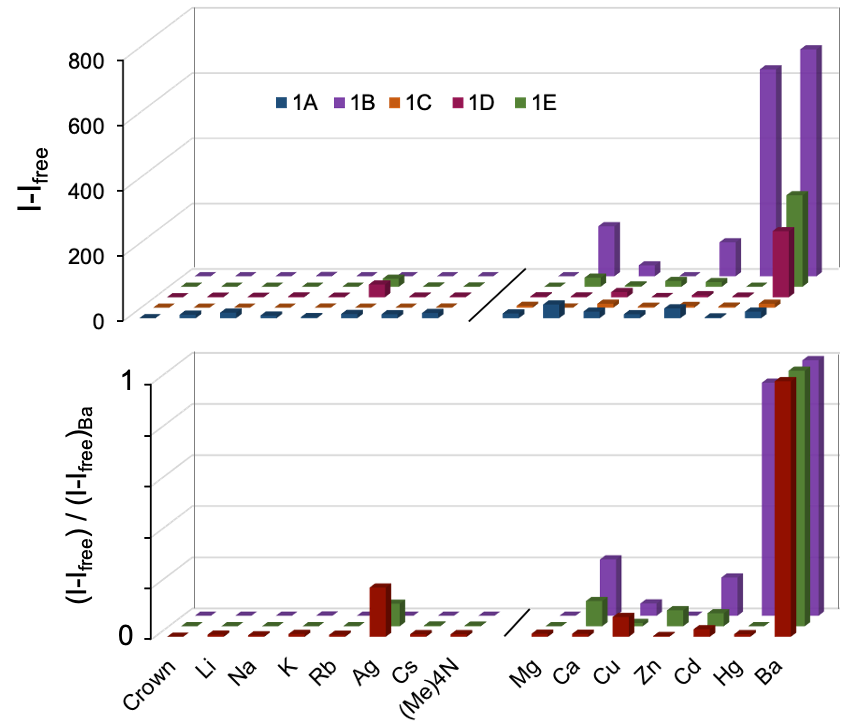}
\caption{Top: ion perchlorate (5$\times$10$^{-6}$~M) response to \textbf{1A-1E} (1$\times$10$^{-6}$~M) in MeCN solution. Bottom: fluorescence response of \textbf{1B}, \textbf{1D}, and \textbf{1E} with different cation solutions normalized to fluorescence intensity for barium perchlorate.}
\label{fig:Selectivity}
\end{figure}

Fluorescent response to cationic analytes was determined for each of the synthesized probes (1.0 $\mu$M in acetonitrile). Stock solutions of perchlorate salts were prepared in 9:1 mixture of MeCN/H$_2$O to achieve complete dissolution. The final concentration of perchlorate solution was maintained at 5 $\mu$M to achieve 1:5 concentration ratio of probe to ion analyte. The selectivity studies are summarized in \textbf{Fig.~\ref{fig:Selectivity}}, top panel with emission at 530 nm. Molecular probes \textbf{1A} and \textbf{1C} are not effective optical sensors for almost all tested ions. A small two-fold response was found in the case of calcium with probe \textbf{1A}. In contrast, \textbf{1B} showed excellent sensitivity towards Ba$^{2+}$ and Hg$^{2+}$ in addition to a lesser response to Ca$^{2+}$. In the presence of Ba$^{2+}$ and Hg$^{2+}$ solutions, ca. 30-fold increases in fluorescence intensities were observed. Ca$^{2+}$ and Cd$^{2+}$ ions also showed eight-fold and five-fold increase, respectively for probe \textbf{1B}.

\begin{figure}[t]
\centering
\includegraphics[width=0.9\linewidth]{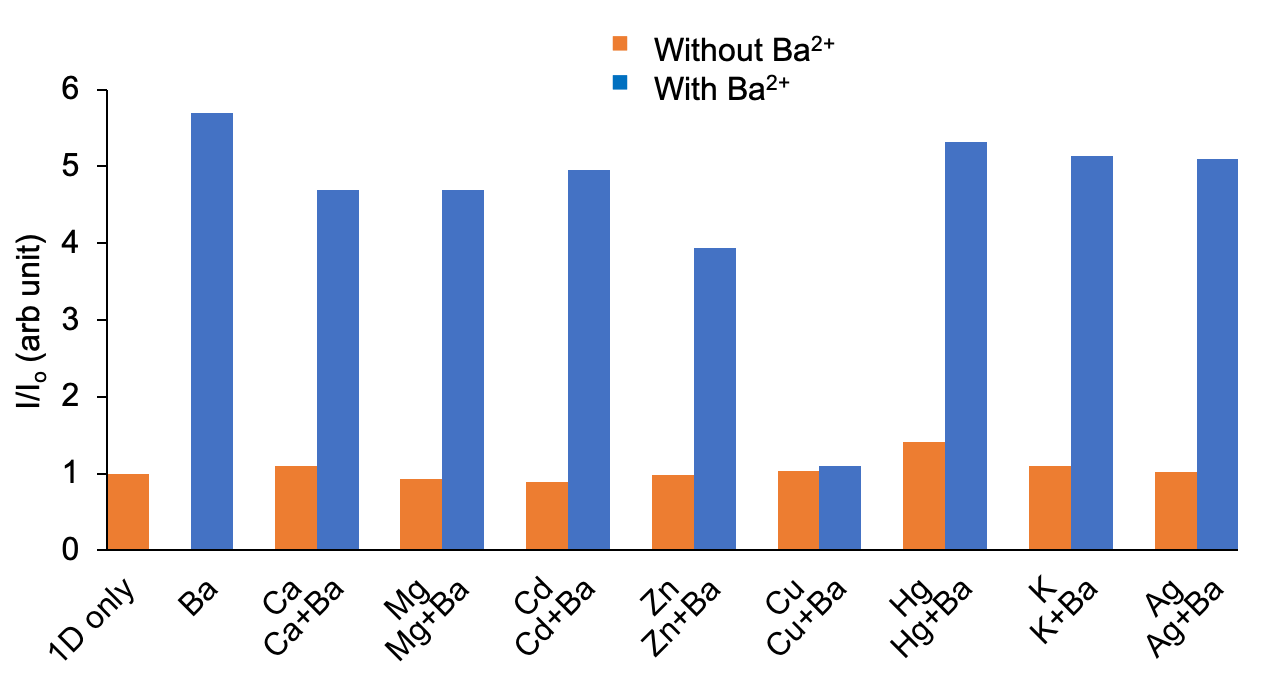}]
\caption{Fluorescence response of \textbf{1D} (2$\times$10$^{-6}$~M) to Ba$^{2+}$ (5$\times$10$^{-6}$~M) in the presence of other metal ions (Ions, 5$\times$10$^{-6}$~M) in acetonitrile. Slit width 2.5/5.0.}
\label{fig:normalizedselecteddata}
\end{figure}

\begin{figure*}[t]
\centering
\includegraphics[width=0.95\linewidth]{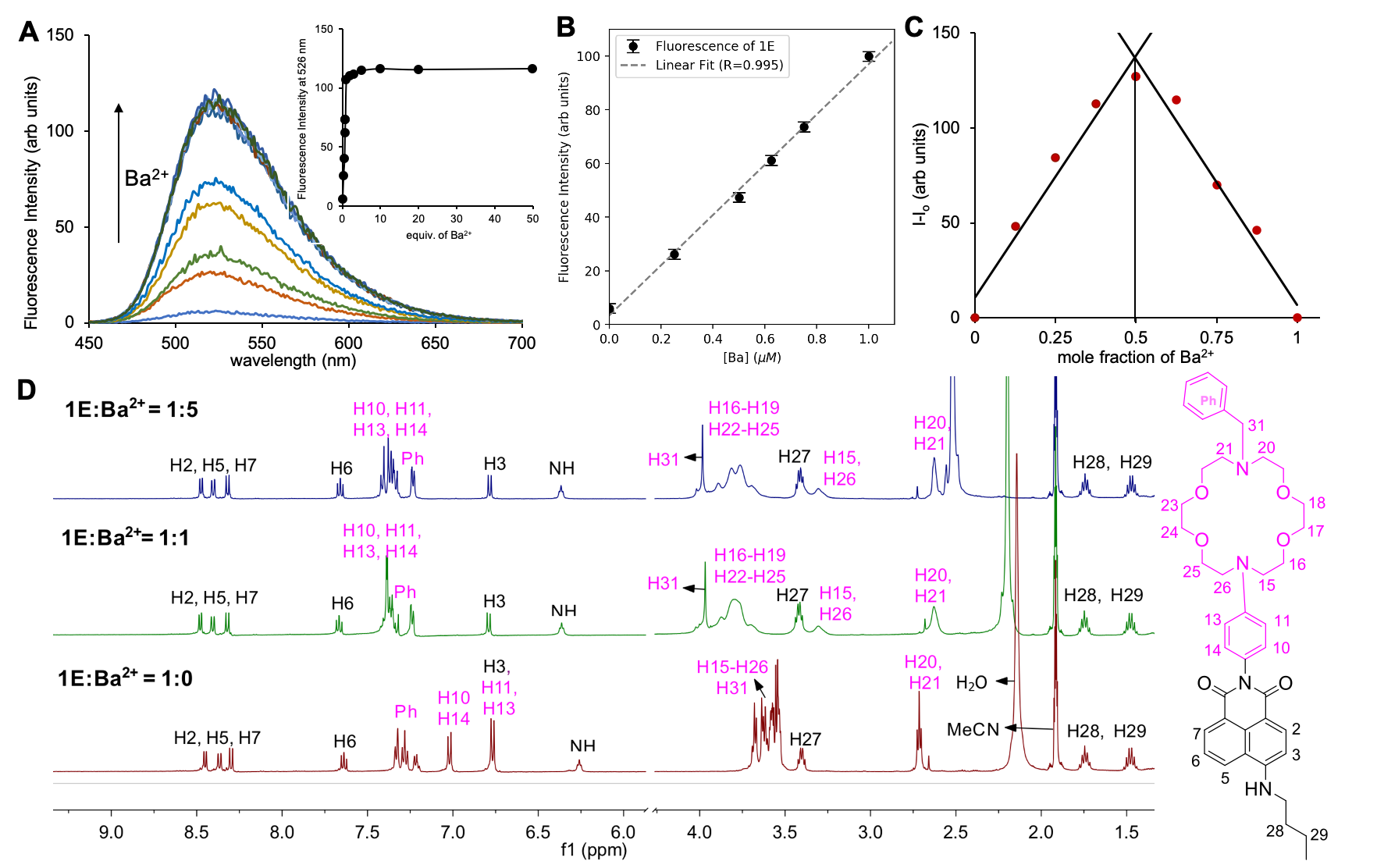}
\caption{\textbf{A}: Fluorescence titration spectra of \textbf{1E} (1$\times$10$^{-6}$~M) upon addition of \textbf{Ba}$^{2+}$ (0-50$\times$10$^{-6}$~M) in acetonitrile ($\lambda_{ex}$ = 430 nm). \textbf{B}: Linearity of fluorescence intensity at 526 nm for \textbf{1E} (1$\times$10$^{-6}$~M) in acetonitrile as a function of the concentration of \textbf{Ba}$^{2+}$. \textbf{C}: Job’s plot figure for \textbf{1E-Ba$^{2+}$} at constant total concentration 20$\times$10$^{-6}$ M of \textbf{1E} and barium in acetonitrile.\textbf{D}: $^1$H NMR titration spectra of probe \textbf{1E} with the addition of 0.0 (red trace), 1.0 (green trace), and 5.0 (blue trace) equiv. of Ba$^{2+}$ in acetonitrile-d$_3$. Inset shows the plot of fluorescence change as a function of \textbf{Ba}$^{2+}$ concentration. Slit width 2.5/5.0 }
\label{fig:Titration}
\end{figure*}

Two novel 1,8-naphthalimide derivatives containing diazacrown ether binding domains \textbf{1D} and \textbf{1E} were found to have excellent selectivity to Ba$^{2+}$. \textbf{1D} and \textbf{1E} showed solution phase emission enhancements of ~11-fold and 22-fold, respectively, in the presence of Ba$^{2+}$. Among different metal ions (Mg$^{2+}$, Ca$^{2+}$, Cu$^{2+}$, Zn$^{2+}$, Cd$^{2+}$, Hg$^{2+}$, Ba$^{2+}$, Li$^+$, Na$^+$, K$^+$, Rb$^+$, Ag$^+$, Cs$^+$, and (CH$_3$)$_4$N$^+$) studied, only Ba$^{2+}$ showed significant increased fluorescence for probe \textbf{1D} and \textbf{1E}. Other metal ions showed little to none (<2$\times$) fluorescence intensity. Normalized fluorescence responses highlight selectivity by ignoring differences in absolute brightness, and show \textbf{1D} and \textbf{1E} with highest selectivity For Ba$^{2+}$ (\textbf{Fig.~\ref{fig:Selectivity}}, bottom panel). Barium sensitivity remained in competitive fluorescence experiments (\textbf{Fig.~\ref{fig:normalizedselecteddata}}). The barium selectivity of probe \textbf{1D} is observed in all cases, except with copper, where direct probe oxidation is likely. Combined, these results are clear evidence of increased selectivity of two novel optical sensors \textbf{1D }and \textbf{1E} to Ba$^{2+}$.

\subsection{Fluorescence titration and binding studies}
Fluorescence titration curves presented in \textbf{Fig.~\ref{fig:Titration}A}, \textbf{SI Fig. S5A}, and \textbf{SI Fig. S6A} show that the emission maxima are obtained before five equivalents of Ba$^{2+}$ are added for probes \textbf{1E}, \textbf{1B}, and \textbf{1D} respectively. Titration studies performed with probes \textbf{1B}, \textbf{1D}, and \textbf{1E} with an increasing concentration of barium perchlorate in acetonitrile gave dissociation constant values (k$_d$) of 0.72$\times$10$^{-6}$~M, 0.59$\times$10$^{-6}$~M and 0.75$\times$10$^{-6}$~M, respectively. These low dissociation constants reveal effective binding between each probe and Ba$^{2+}$.  Solution phase limits of detection (LOD) were calculated using linear fitting curves of fluorescence titration data and the formula LOD = 3$\sigma$/k, where $\sigma$ is the standard deviation of the blank sample without barium and k is the slope between intensity versus barium concentration,\cite{wang2010selective} resulting in 0.0288$\times$10$^{-6}$~M, 0.074$\times$10$^{-6}$~M, and 0.006$\times$10$^{-6}$~M detection limits for probe \textbf{1E}, \textbf{1B}, and \textbf{1D}, respectively. \textbf{Fig.~\ref{fig:Titration}B}, \textbf{SI Fig. S5B}, and \textbf{SI Fig. S6B } show good linear correlation for probes \textbf{1E} (R$^2$ = 99.5), \textbf{1B} (R$^2$ = 98.5), and \textbf{1D} (R$^2$ = 99.1), respectively.
Overall results show that the probe \textbf{1B}, \textbf{1D} and \textbf{1E} can detect Ba$^{2+}$ in solution at nanomolar concentrations and therefore, can be highly practical fluorescent sensors for selective detection of Ba$^{2+}$.
The stoichiometry of binding for molecular probes \textbf{1B} and \textbf{1E} with barium were calculated by Job’s method. The results, presented in supplementary \textbf{SI Fig. S5C} and  \textbf{Fig.~\ref{fig:Titration}C}, show the inflection point of two linear fitting curves at 0.46 and 0.50 mole fraction for complex \textbf{1B-Ba$^{2+}$} and \textbf{1E-Ba$^{2+}$} respectively. This indicates a 1:1 binding stoichiometry between probes \textbf{1B} and \textbf{1E} with Ba$^{2+}$.  Additionally, complex \textbf{1B-Ba$^{2+}$} was observable in high resolution mass spectrometric (HRMS) analysis of solution of \textbf{1B} and barium perchlorate (\textbf{SI Fig. S5D}).

\subsection{$^1$H NMR studies}
Complexation between Ba$^{2+}$ and fluorescent probes \textbf{1B}, \textbf{1D}, and \textbf{1E} were further demonstrated by $^1$H NMR experiments performed with 1:1 and 1:5 molar ratio of probe and Ba(ClO$_4$)$_2$ in acetonitrile-\textit{d}$_3$. Ba$^{2+}$ addition resulted in downfield shifts of \textit{N}-phenyl-aza crown ether protons, while the chemical shift values of protons in naphthalimide fluorophore group have minimal or no change (\textbf{Fig.~\ref{fig:Titration}D}). Noticeably, an upfield shift in $\alpha$-protons of the anilino nitrogen in azacrown ether was also observed (H15 and H26 in \textbf{Fig.~\ref{fig:Titration}D}) in all molecular probes (for \textbf{1B} and \textbf{1D}, see \textbf{SI Fig. S7}), suggesting that the nitrogen atom of \textit{N}-phenyl-aza crown ether is not directly involved in ion binding. Similarly, the effect of Ba$^{2+}$ on N-benzyl group in probe \textbf{1E} was observed, as seen with downfield shifts of both aromatic and benzylic protons \textbf{(Fig.~\ref{fig:Titration}D)}. Additionally, in fluorescent probe \textbf{1D} and \textbf{1E}, all $\alpha$-protons of two nitrogen atoms in the diazacrown ether structure showed upfield shift indicating no interaction between both nitrogen atom of diazacrown and Ba$^{2+}$. The results obtained from $^1$H NMR experiments with probes clearly show the binding of Ba$^{2+}$ on the receptor crown ether units with no significant binding on naphthalimide fluorophore units. 
Comparative $^1$H NMR studies were carried out with Ba$^{2+}$, Hg$^{2+}$and K$^+$ in the case of probe \textbf{1B}, which show distinct features in the NMR spectra. With Hg$^{2+}$ addition, all monoazacrown methylene protons were deshielded, indicating binding of nitrogen atom of {N}-phenyl-aza crown ether with Hg$^{2+}$ (\textbf{SI Fig. S8}). With K$^+$, however, no observable change was found in the chemical shift values of the protons of both naphthalimide and azacrown ether moieties (\textbf{SI Fig. S9}). 

\begin{figure}[t]
\centering
\includegraphics[width=0.99\linewidth]{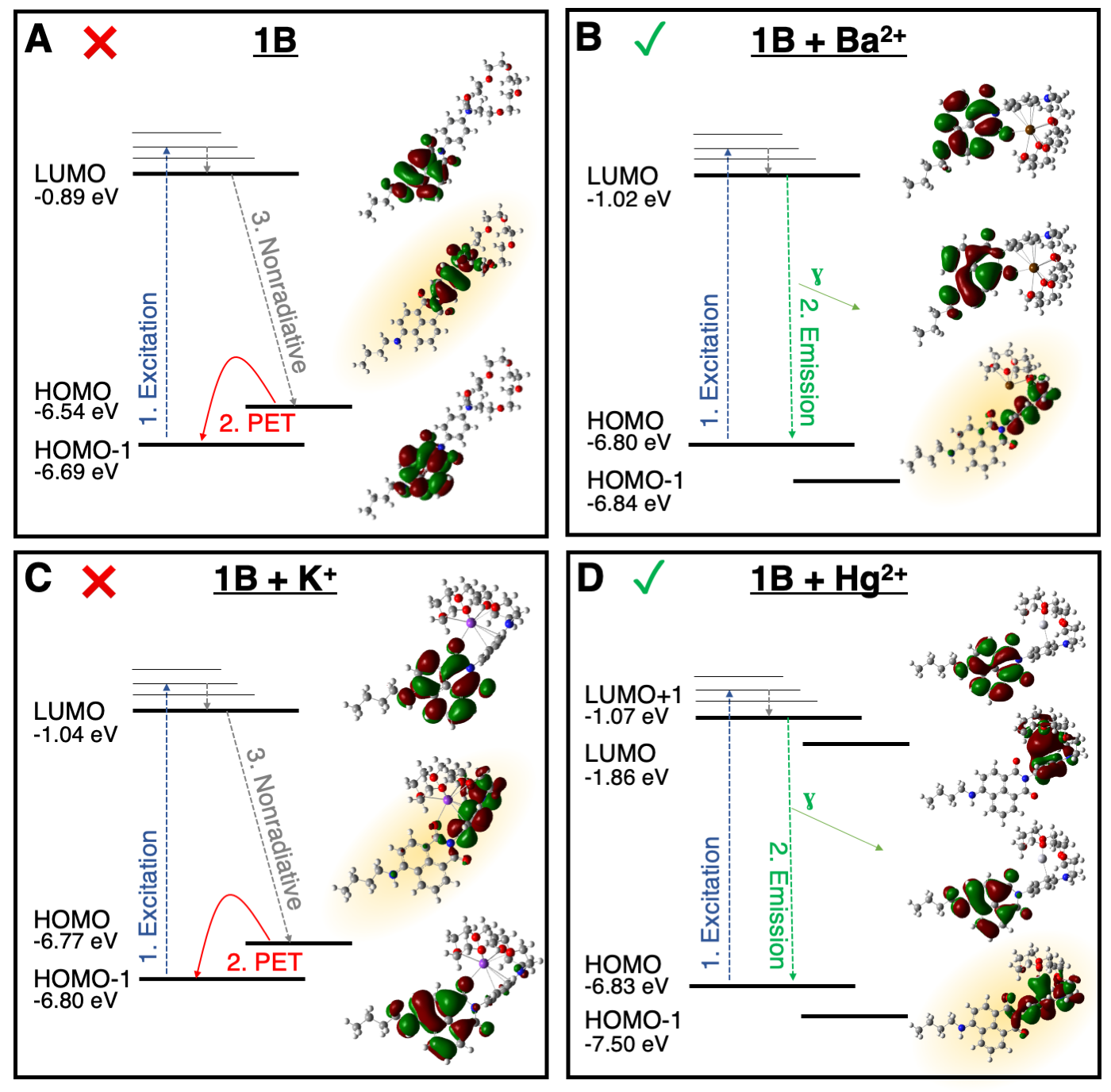}
\caption{Simulated orbitals of \textbf{1B}.  The mechanism of PET and CHEF is illustrated by the change in energy of the nitrogen-dominated orbital, highlighted, upon binding various cations.}
\label{fig:Compute1B}
\end{figure}

\begin{figure}[t]
\centering
\includegraphics[width=0.99\linewidth]{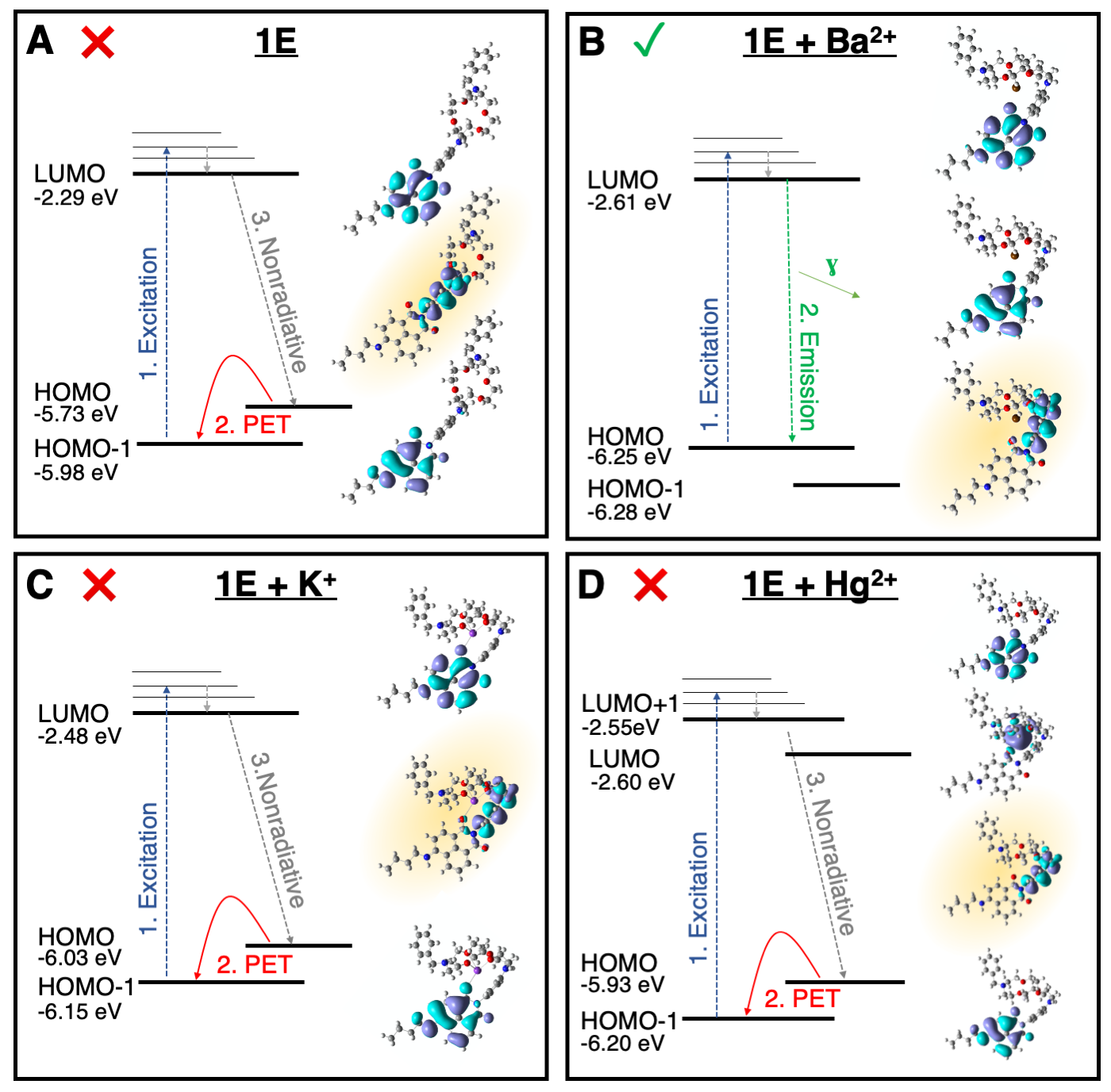}
\caption{Simulated orbitals of \textbf{1E}.  Enhanced Ba$^2+$ selectivity is illustrated by comparing changes in energy of the nitrogen-dominated orbital, highlighted.}
\label{fig:Compute1E}
\end{figure}

\subsection{Computational studies} 

Computational models were studied using TDDFT to predict the fluorescence behaviour of sensors in response to ions in acetonitrile solution.  As a test suite we considered molecules \textbf{1B} and \textbf{1E}, both unchelated and chelated with Ba$^{2+}$, K$^+$, and Hg$^{2+}$.  The results support the proposed PET-mediated off-state within unchelated mono- and diazacrown ether naphthalimides, which is turned-on by chelation-enhanced fluorescence (CHEF) upon ion binding. In the unbound \textbf{1B} species (\textbf{Fig.~\ref{fig:Compute1B}}A), excitation of the naphthalimide fluorophore is governed by a HOMO-1 to LUMO transition based on orbital analysis. The electrons in the interstitial HOMO$_{free}$ orbital of the receptor (highlighted) are localized around the \textit{N}-aryl unit and energetically well-positioned to quench the excited state before fluorescence transition can occur from LUMO to HOMO(-1), by a traditional PET mechanism. Upon barium chelation, however, the HOMO$_{free}$ level is drastically stabilized by mixing with Ba$^{2+}$ and becomes the new HOMO-1 (Fig \ref{fig:Compute1B}B), allowing the fluorophore-centered HOMO and LUMO levels of the barium bound species to participate in fluorescence without PET-quenching.

As \textbf{1B} responds to Hg$^{2+}$, but not to K$^+$, we examined the arrangement of frontier molecular orbitals when complexed to these cations. Chelation is effective and lowers the critical PET-enabling HOMO$_{free}$ orbital through binding to both ions. However, in the case of K$^+$, PET quenching is still possible (Fig \ref{fig:Compute1B}C), predicting a non-fluorescence response to K$^+$ consistent with experiment.  Hg$^{2+}$ chelation results in notable lowering of the HOMO, LUMO, and LUMO+1 energy states relative to the free chemosensor(Fig \ref{fig:Compute1B}D); the LUMO and LUMO+1 states show substantial lowering in energy including a reordering of states, not unusual among heavy atom binding.\cite{lee2012mechanism} Excitation of the fluorophore-centered electrons (HOMO to LUMO+1) are not quenched by PET. Thus CHEF occurs with Hg$^{2+}$ addition to \textbf{1B}, similar to the barium chelation, albeit with a different rearrangement of orbital energies.

Analysis of the selective diazacrown ether chemosensor \textbf{1E} showed similarity to \textbf{1B} when free and bound to Ba$^{2+}$ and K$^+$, especially with respect to the critical relative location of the HOMO$_{free}$ orbital (highlighted in \textbf{Fig.~\ref{fig:Compute1E}}). However, Hg$^{2+}$ binding resulted in a weaker effect on the HOMO$_{free}$ orbital, and the fluorescence of Hg$^{2+}$ bound species is still quenched by internal PET from HOMO to HOMO-1. Overall, the theoretical studies support experimental results for the enhanced selectivity of \textbf{1E} to the Ba$^{2+}$.

In conclusion, the theoretical models correctly predict the observed switch-on fluorescence patterns across all ions and species tested. These are summarized in \textbf{Table~\ref{table:computationsummary}}.

\begin{table}[h]
\centering
\footnotesize
\begin{tabular}{|c| c| c c |}
\hline
 &  & Fluorescence & Fluorescence \\
 Probe & Ion & predicted & observed \\
\hline
\hline
\textbf{1B} & -- & No & No \\
\textbf{1B} & Ba$^{2+}$& Yes & Yes\\
\textbf{1B} & K$^{+}$& No & No\\
\textbf{1B} & Hg$^{2+}$ &Yes & Yes \\
\hline
\textbf{1E} & -- & No & No \\
\textbf{1E} & Ba$^{2+}$& Yes & Yes\\
\textbf{1E} & K$^{+}$& No & No\\
\textbf{1E} & Hg$^{2+}$ &No & No \\
\hline
\end{tabular}
\centering

\caption{Summary of computational results compared with experimental observations of switch-on fluorescence.}
\label{table:computationsummary}
\end{table}

\begin{figure*}[t]
\centering
\includegraphics[width=0.99\linewidth]{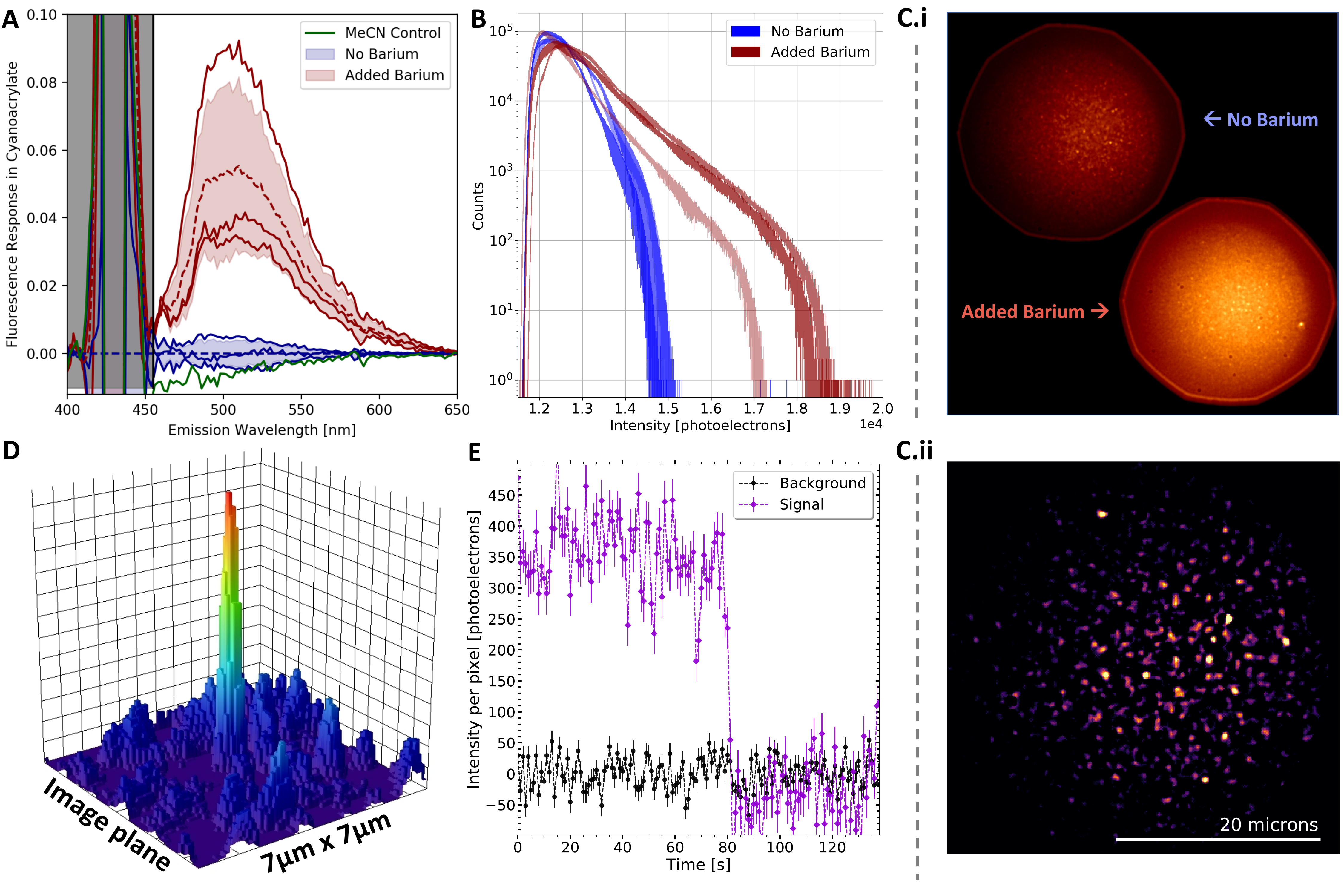}
\caption{Solid-matrix measurements of barium sensing fluorophore \textbf{1D}. \textbf{A}: bulk spectrophotometry in dried cyanoacrylate matrix; \textbf{B}: Raw pixel histogram of three \textbf{1D} coated slides imaged at the single molecule level; \textbf{C}: Single molecule level microscope images of the \textbf{1D} layer with and without Ba$^{2+}$ \textbf{(i)} and with Ba$^{2+}$ after background removal \textbf{(ii)};  \textbf{D}: A specific single reconstructed barium-chelated \textbf{1D} molecule; \textbf{E}: Barium-chelated \textbf{1D} molecule fluorescence time trajectory showing single-step photo-bleaching characteristic of SMFI detection.}
\label{fig:SingleMolFigs}
\end{figure*}

\subsection{Dry SMFI of Ba$^{2+}$} 

To demonstrate single ion sensing in a dry environment, compound \textbf{1D} was suspended in cyanoacrylate polymer. Polyvinylacetate and polystyrene matrices were initially investigated with relatively poor solvochromatic effects on bulk fluorescence.  Barium-induced fluorescence was first verified by spectrophotometry in bulk matrix. A strong fluorescent response was observed across all three tested locations on three slides, though with significant variability observed due to imperfect uniformity of layer deposition.  \textbf{Fig.~\ref{fig:SingleMolFigs}A} shows the background-subtracted response of slides with no added Ba$^{2+}$, and after Ba$^{2+}$ wash.  Fluorescence is normalized to activity on the rising edge of the excitation peak before background subtraction.  Mean and standard deviation are shown as dashed and filled regions, in addition to individual response curves. Also shown overlaid is a control measurement using a pure solvent wash--resulting in slightly reduced fluorescence, likely due to removal of some ionic contamination or fluorescent emitters from the surface. These data demonstrate that fluorescent response to Ba$^{2+}$ is maintained in the dry phase, as observed in our past work with molecules of this family~\cite{thapa2019barium}.

The imaging modality for SMFI in this paper is air-coupled epifluorescent microscopy.  This is distinct from prior work~\cite{mcdonald2018demonstration} on SMFI imaging of Ba$^{2+}$ in solution, which used Total Internal Reflection Fluorescence (TIRF).  TIRF offers combined benefits of reduced background through by exciting only at the glass-sample interface; and oil couplings that allow for high numerical aperture imaging.  The oil required for through-objective TIRF, however, is likely problematic within the high purity xenon environment of our target application.  While the oil-free imaging modality used here introduces additional practical challenges,  realization of single Ba$^{2+}$ imaging in this way encapsulates an important practical step toward  application in time projection chambers.


A set of slides was prepared for single-ion level imaging of barium, with compound \textbf{1D} suspended at 10~pM concentration in the cyanoacrylate matrix. Slides were washed with a 1~nM Ba$^{2+}$ solution.  A stark increase in the net fluorescence intensity of the sample was observed upon addition of Ba$^{2+}$.  Three slides were tested, with image sequences taken across three locations both before and after the barium solution was introduced. The increase in intensity was quantified via raw pixel histogram for each slide, shown in \textbf{Fig. \ref{fig:SingleMolFigs} B}.  A robust increase in bright pixels was observed in all cases, indicative of turn-on fluorescence.

The detected fluorescence in these images originates from molecules at various distances from the focal plane, with near-focus points appearing very bright, and out-of-plane candidates dimmer. Both before and after addition of Ba$^{2+}$, an array of spots of various brightness can be observed, undergoing discrete photo-bleaching transitions over time, with the number dramatically enhanced in the barium-added images.  An example of a before/after comparison is shown in \textbf{Fig. \ref{fig:SingleMolFigs} C.i}, where in both cases, bright points corresponding to distinct barium-\textbf{1D} complexes can be visually identified.  

Using a sample with 10~pM barium perchlorate solution applied to the 10~pM sensor matrix, the interpretation of these bright emitters as single molecules was confirmed.   \textbf{Fig. \ref{fig:SingleMolFigs} C.ii} shows an example image processed and filtered according to the algorithm described in the methods section, which is used as input to time-series analysis.  The well localized candidate spots in this image, an example of which is shown isolated in \textbf{Fig. \ref{fig:SingleMolFigs} D}, are observed to exhibit discrete photo-bleaching transitions, as shown in \textbf{Fig. \ref{fig:SingleMolFigs} E}.  This behaviour, also observed visually for the bright spots in the denser samples, as well in previous work in solution phase~\cite{mcdonald2018demonstration}, confirms the single molecule interpretation of these fluorescent emitters.


\subsection{Toward Ba$^{2+}$ tagging in Xenon}

An aspect that remains undemonstrated here is the capture of Ba$^{2+}$ within an environment of high pressure xenon gas.  Our past theoretical work~\cite{Bainglass:2018odn} demonstrated that solvation-like effects with xenon are expected, with the shell configuration depending on gas temperature and pressure.  The xenon shell typically contains 7-10 xenon atoms for Ba$^{2+}$ at 10-15~bar. Capture by a molecular layer must free Ba$^{2+}$ from this shell of accompanying spectator atoms, typically held at a binding energy of 3-4 eV.  The Ba$^{2+}$ binding energy of the molecules developed in this work is calculated to be around 0.2~eV in MeCN. However, calculations of the binding affinity in the gas phase, removing polarization effects of the solvent, show a much enhanced binding energy of -10.5~eV. This is sufficient to extract the target Ba$^{2+}$ from its weakly-attached xenon neighbors, so efficient capture of Ba$^{2+}$ from high-pressure xenon gas is expected.   Notably, the xenon solvation shell may also offer useful protection until  proximity adequate for capture occurs.

\begin{figure}[t]
\centering
\includegraphics[width=0.99\linewidth]{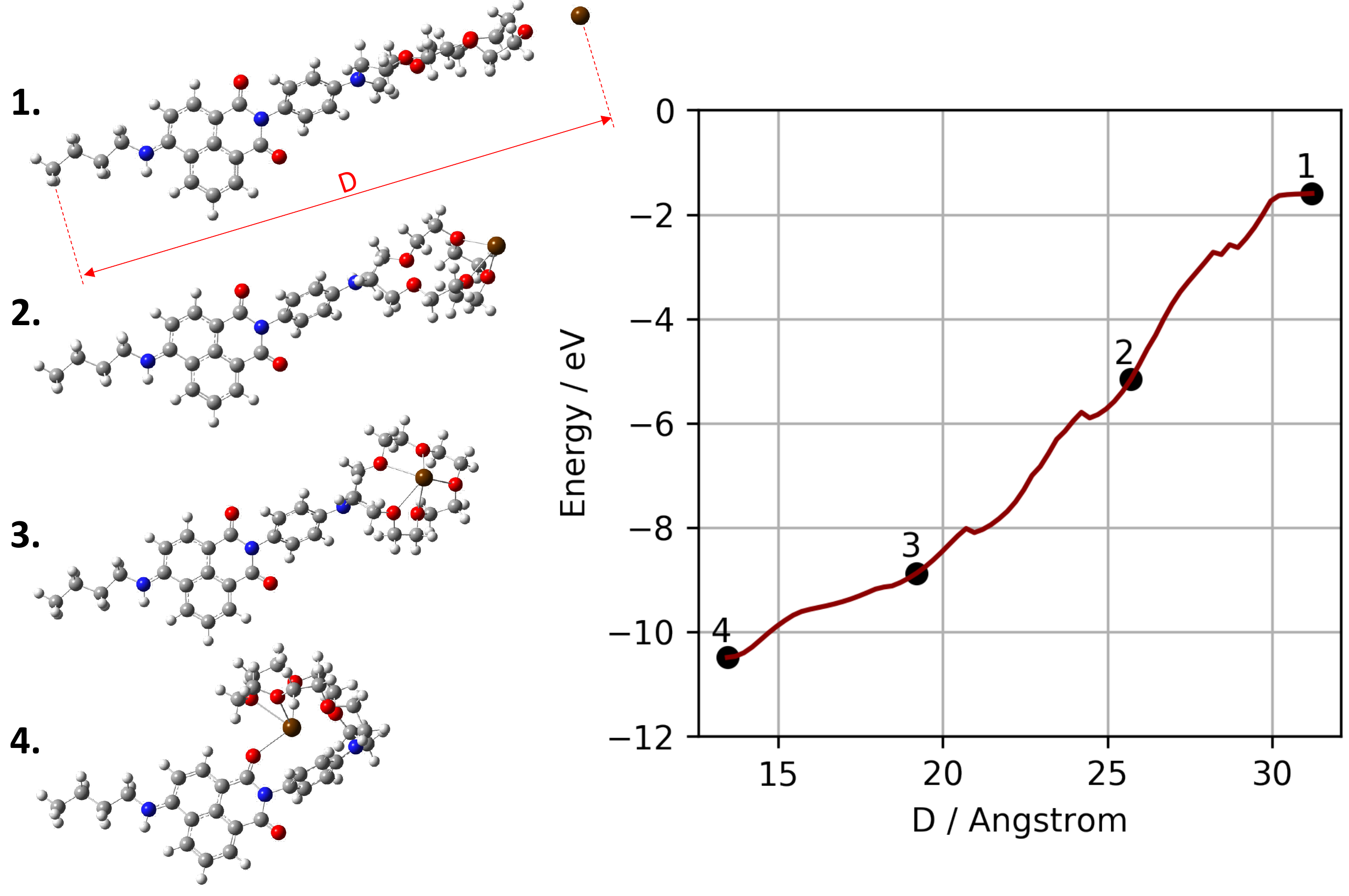}
\caption{Simulation of ionic capture (left), and potential energy surface in the gas phase as a function of ion-to-surface distance (right)}
\label{fig:CaptureSim}
\end{figure}

A pertinent question for design of a molecular sensing layer at the gas-solid interface is the optimal surface density of probe molecules.  An ultra-dense fluorophore layer is likely to suffer from collective quenching effects, whereas an overly sparse one may not efficiently capture arriving ions.  To this end we undertook computations of the range of ion capture.  Complexation can be considered to be inevitable when the ion-molecule proximity is such that the binding energy is more than a few times the thermal energy (kT$\sim 0.02$~eV at STP). To evaluate the effective capture range, the most stable geometry of the complex \textbf{1B-Ba$^{2+}$} was evaluated as a function of ion-to-surface distance in gas phase (\textbf{Fig. \ref{fig:CaptureSim}}). Upon allowing Ba$^{2+}$ to move closer to the azacrown ether surface, a bent geometry was obtained where the ion is bound to naphthalimide oxygen along with four ethereal oxygen of the azacrown ether substructure (structure 4). This bent geometry of the complex is consistent with the NMR titration study in solution, in which the nitrogen atom of the azacrown ether did not appear to directly bind with the electron-withdrawing Ba$^{2+}$ ion.

Beyond a range of 32 Angstroms from the anchor point, our simulations fail to converge. The binding energy at this distance is still large, at 1.75~eV.  Extrapolating the trend observed in the potential energy surface, it appears that the effective binding range of the molecule is somewhere between 32 and 40 Angstroms.  This suggests a range of densities for future monolayer construction, in order to realize efficient binding and reduce intermolecular distortions to fluorescent events. Exploration of the power of sensitive, semi-dense monolayers for ion capture at the solid-gas interface is the next immediate step in this ongoing program.

\section{Conclusion}
We have designed, synthesized, and studied barium selective fluorescent sensors to image Ba$^{2+}$ via SMFI in dry environments. We found that a new class of visible-spectrum 1,8-naphthalimide derivatives with diaza-18-crown-6 substituents dramatically improved the selectivity of the sensors to Ba$^{2+}$ without significantly compromising sensitivity. Solution-phase fluorescence and NMR experiments supported a PET mechanism enabling turn-on fluorescence sensing in the presence of metal ions. Experiments showed strong binding of Ba$^{2+}$ to the sensor with 1:1 stoichiometry and a nanomolar limit of detection.   Experimentally validated theoretical calculations illuminated the mechanism of fluorescence sensing, in addition to providing insights into expected behaviour as gas-solid interfaces.

SMFI microscopy in air-coupled epifluorescent mode was employed for sensing of single Ba$^{2+}$ ions with these fluorophores. Cyanoacrylate was found to be an effective support medium for the aforementioned fluorophores, and due to the sensor's strong selectivity for barium, no competitive binding agent was required to achieve single molecule sensitivity with these molecules.  Single Ba$^{2+}$ candidates were resolved spatially and identified through single-step photo-bleaching transitions, with enhanced prevalence in barium-washed samples.  Realization of single Ba$^{2+}$ sensitivity  under these conditions represents an important step toward practical application of this technique within time projection chambers. Such a technique could enable new precision and robustness in searches for $0\nu\beta\beta$ in xenon gas.

\subsection{Acknowledgements}
This work was undertaken as part of an ongoing interdisciplinary program to develop barium tagging technologies for the NEXT experiment.  We gratefully acknowledge support from the Department of Energy under awards DE-SC0019054 and DE-SC0019223, and from the University of Texas at Arlington.  NMR experiments were made possible by support from the NSF, CHE-0840509 (CRIF: MU).  Mass spectrometry experiments were completed in the Shimadzu Center for Analytical Chemistry at UT Arlington.

\newpage

\bibliography{paperdraft}

\end{document}